\documentclass[aps, twocolumn,  showpacs]{revtex4}

\begin{document}
\title{Comment on 'New perspective on thermodynamics of spacetime: The emergence of unimodular gravity and the equivalence of entropies'}
\author{S. C. Tiwari \\
Department of Physics, Institute of Science,  Banaras Hindu University,  and Institute of Natural Philosophy, \\
Varanasi 221005, India }
\begin{abstract}
In this comment on the article by Alonso-Serrano and Liska (arXiv: 2008.04805) a formal resemblance between their expression of curvature scale and the scale dependent on the matter energy-momentum ambiguity of Finkelstein et al (JMP, 2001) is pointed out. Physical significance of this observation is also discussed.
\end{abstract}
\maketitle

The role of covariant divergence law for the energy-momentum tensor in general relativity has varied perceptions regarding its physical interpretation. In unimodular relativity it has been shown \cite{1} that the cosmological constant appears as an integration constant. Local matter energy conservation acquires added importance in this theory \cite{2}. Finkelstein et al \cite{3} re-examined the issue raised in \cite{2} presenting a thorough discussion on the variational principle for the infinitesimal variations of the conformal metric tensor field. However, the conclusion of the authors that the energy-momentum tensor satisfies the standard covariant divergence law and that the cosmological constant is a constant of integration was questioned in my note \cite{4}. In a recent paper, Alonso-Serrano and Liska \cite{5} show that unimodular relativity seems more natural than general relativity in the thermodynamic derivation of Einstein field equation. I consider this to be a welcome result in the light of my earlier work \cite{6}. However the problem of local energy conservation is not addressed satisfactorily in \cite{5}. The present comment has origin in a remarkable formal similarity between the expression of the curvature scale $\lambda$ in \cite{5} and the expression (9) obtained in \cite{4} that depends on the ambiguity in the energy-momentum tensor introduced by Finklestein et al \cite{3}. Is it accidental? Or, does it involve deep physical implication?

Let us first establish the claimed formal resemblance. In the notation of \cite{3} the Einstein field equation given by Eq.(4) in \cite{4} is
\begin{equation}
 G^{\mu\nu} - \frac{\lambda_F}{2} g^{\mu\nu} = 8 \pi G T^{\prime \mu\nu}
\end{equation}
where 
\begin{equation}
 T^{\prime \mu\nu} = T^{\mu\nu} +\frac{1}{\sqrt{-g}} \frac{\delta \sqrt{-g} \Delta_M L}{\delta g_{\mu\nu}}
\end{equation}
We have changed $\lambda$ to $\lambda_F$ in Eq.(1) and light speed is set equal to unity. The assumed ambiguity is 
\begin{equation}
 \Delta_M L = (\frac{\mu(x)}{\sqrt{-g}} - 1) l_M
\end{equation}
Here $l_M$ is a function of matter field and $g_{\mu\nu}$, and fundamental measure $\mu(x)$ defines the unimodular condition
\begin{equation}
 \sqrt{-g}~ d^4x = \mu(x) ~d^4x
\end{equation}
It is straightforward to calculate the ambiguity using Eq.(3)
\begin{equation}
 \Delta T^{\mu\nu} = \frac{g^{\mu\nu} ~ l_M}{2}
\end{equation}
Taking the trace of Eq.(1) we get
\begin{equation}
 2 \lambda_F = -R - 8 \pi G (T + 2 l_M)
\end{equation}
In \cite{3} the ambiguous term (5) has a negative sign; it appears there is a lack of clarity whether $\Delta T^{\mu\nu}$ is equal to $(T^{\prime \mu\nu} - T^{\mu\nu})$ or $(T^{ \mu\nu} - T^{\prime \mu\nu})$. Using expression (24)  of \cite{3} the expression for $\lambda_F$ is different than given by Eq.(6); denoting it by $\lambda_F^\prime$ we have
\begin{equation}
 2 \lambda^\prime _F = - R - 8 \pi G (T - 2 l_M)
\end{equation}

Now, in the thermodynamical approach \cite{5} the unimodular field equation is obtained to be 
\begin{equation}
 R^{\mu\nu} - \frac{1}{4} R g^{\mu\nu} = 8 \pi G ( \delta<T^{\mu\nu}> - \frac{1}{4} \delta<T> g^{\mu\nu})
\end{equation}
Here $<T^{\mu\nu}>$ is the expectation value for quantum fields. A length parameter $l$ for geodesic local causal diamond, and a curvature scale $\lambda$ - for maximally symmetric spacetime appearing in $G^{00} = - \lambda g^{00}$ are introduced. Calculation of matter entanglement entropy shows that for nonconformal fields a scalar $X$ is needed that depends on $l$. Trace of the equations of motion determine $\lambda$
\begin{equation}
 \lambda = \frac{R}{4} + 8 \pi G ( \frac{<\delta T>}{4} - \delta X)
\end{equation}

It is easy to verify that formal equivalence of (7) and (9) holds with the following correspondence 
\begin{equation}
 \lambda \rightarrow -\frac{\lambda^\prime_F}{2} ; ~\delta<T> \rightarrow T ; ~\delta X \rightarrow \frac{l_M}{2} 
\end{equation}
First two identifications in (10) have no surprise: compare Einstein equation (1) given here and Eq.(27) in \cite{5}, and also note that quite often semiclassical arguments are used for expectation value for the energy tensor. The correspondence $ \delta X \rightarrow  \frac{l_M}{2}$ is the most intriguing part of the formal equivalence shown here. Let us try to explain it in the following.

In \cite{3} the unimodular ambguity in the action is purely a formal construct, in fact, authors state that, 'No physical result may depend on the choice of $\Delta_M L$ , and so no physical  experiments can determine $\Delta_M L$'. Though in the concluding part of \cite{3} quantum vacuum effects for a variable cosmological constant are admitted, the physical interpretation of the ambiguity remains unexplained. Note that $\delta X$ just like the ambiguity term $l_M$ does not appear in the gravitational field equation, however it has physical interpretation as a measure of the nonconformability of quantum fields \cite{5}. It would be natural to relate  it with some kind of spacetime fluctuations in unimodular ambiguity. Do they have observable effect? Since $\delta X$ in \cite{5} and $l_M$ in \cite{3} do not affect gravity one may have to look for Aharonov-Bohm phase shift like effects, if any.

To summarize we make following points. [1] :- Einstein field equation as an equation of state of thermodynamical spacetime necessitates to probe spacetime structure at a fundamental level. Already in 1971 a cellular structure for spacetime was envisaged for unimodular relativity \cite{1}. [2] :- Authors \cite{5} show the equivalence of Clausius and matter entanglement entropy. A causal diamond filled with conformal matter is argued to have equivalence of two entropies. On the other hand, the unimodular theory in \cite{1,3} is based on the action principle. Could there be a unified picture for both? In this connection de Broglie's speculation \cite{7} that the principle of least action is a particular case of the second law of thermodynamics deserves attention \cite{8}.

To conclude, the new perspective on the thermodynamics of spacetime \cite{5}, speculations on the discrete spacetime manifold in unimodular relativity \cite{1,3}, and statistical nature of spacetime envisaged in \cite{6,8,9} combined together suggest radical revision on the physical conception of  space and time reality at a fundamental level.

\end{document}